# Polarization gaps in spiral photonic crystals

Jeffrey Chi Wai Lee and C.T. Chan*
*Department of Physics,
Hong Kong University of Science and Technology,
Clear Water Bay, Kowloon, Hong Kong, China*

**Abstract**: We studied the optical properties of a dielectric photonic crystal structure with spirals arranged in a hexagonal lattice. The dielectric constant of the material is 9 and the filling ratio is 15.2%. We found that this kind of structure exhibits a significant polarization gap for light incident along the axis of the spirals. The eigenmodes inside the polarization gap are right-hand (left-hand) circularly polarized depending on the whether the spirals are left-handed (right-handed). The transmission spectrum of a slab of such a structure has been calculated and matches well with the analysis of the eigenmodes.

Spiral is a kind of three-dimensional structure that has chiral character. This kind of structure attracts attention because of its optical activity [1] and its geometric resemblance to the diamond structure [2]. Recent studies by [3,4] also showed that spiral-structured photonic crystals process complete photonic bandgaps and they are also candidates for negative refraction materials [5]. A few innovative techniques have successfully fabricated various forms of spiral structures in the micron and sub-micron scale, and examples including glancing angle deposition (GLAD) [4], two-photon processes [6,7], and holography-lithography [8]. We also find by computer simulations that there are at least two ways to produce a periodic array of separated hollow spiral structures (spring-like structure) by the holographic lithography method. Spring-like spiral structures containing metallic nano-Ag can also be made by two-photon writing processes [7]. The spirals fabricated by holographic lithography are separated and hence their properties are expected to be different from those in [3]. The purpose of this paper is to examine the band structure and transmission properties of a periodic array of spirals and show that they possess significant polarization gaps in which one circular polarization is forbidden inside the photonic crystal.

All the band structures in this paper were calculated by the plane-wave expansion method. We used 2000×2000 matrices and doubling the matrix size shows no observable differences. The transmission spectra are calculated by the scattering matrix method SMM[9,10].

The plane wave method was first used to study the spiral structure of Toader et. al. [3] and our band structure results how a perfect match with those in [3]. The spiral structure (with $\varepsilon=11.9$) fabricated by the GLAD method exhibits robust photonic gaps, but we what to know whether the spiral geometry can result in a polarization gap, meaning that it permits the transmission of one circular polarization but rejects the other handedness in some frequencies. The transmission calculation for light incident along the $\Gamma$ to Z direction for circularly polarized light for the Toader structure is shown in Fig. 1. Although the basic building block of their structure is spiral, the transmittance for right and left circularly polarized light are almost the same. The weak chiral character is probably because the spirals are strongly overlapping.

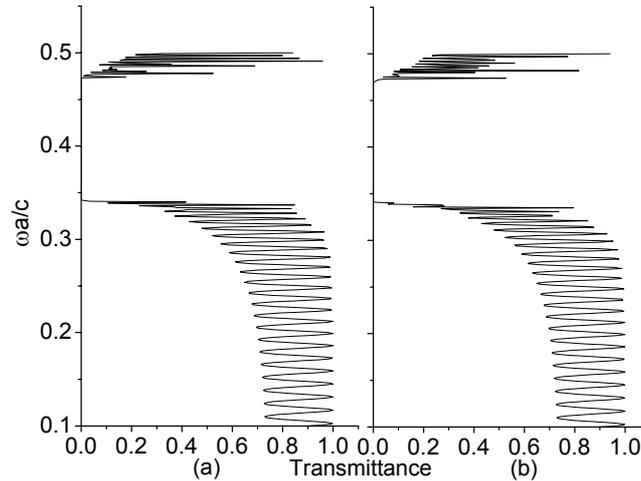

Fig. 1. Transmission spectra of the dielectric structure proposed by Toader [3] for (a) left-hand and (b) right-hand circularly polarized incident plane wave. The slab has 16 unit cells along the Z direction.

We found that well-separated spirals exhibit much stronger chiral character. Such geometrical structures can in fact be made using holographic lithography techniques. For example, a 7-beam configuration with 6 equally-spaced circumpolar linear polarized side beams and a circular polarized central beam will interfere to form spirals that are separated and form a hexagonal periodic array in the x-y plane. This class of structure has been realized experimentally [11]. The band structure was then calculated for a structure consisting of right-hand spirals shown in Fig.2, which is an idealized structure derived from the one made by 7 beam laser interference with one of them being circularly polarized. We focus on the ΓA direction which is parallel to the spiral axis (equivalent to the ΓZ direction in the previous case). The major difference between the one shown in Fig. 2 and the one used in [3] is the connectedness between the spirals. Spirals in the structure shown in Fig. 2 are separated and will a smaller the filling ratio of 15.2% than the one considered in [3].

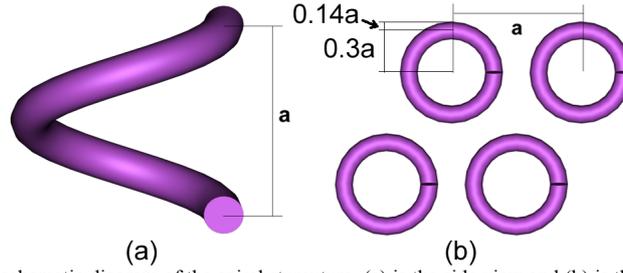

Fig. 2. A schematic diagram of the spiral sturcture. (a) is the side view and (b) is the top view of the array. The ΓA direction is the z axis. The spiral has a dielectric constant ε = 9. The filling ratio of the structure is 15.17 %.

The band structure of the spiral structure and the corresponding transmission spectrum through a stack of 16 layers are shown in Fig. 3. For the SMM calculation, each unit cell is divided into 32 sub-layers in the z direction. We found a sizable polarization gap in the band structure (Fig. 3c) and the polarization gap width to the middle gap ratio is 26.40%. Experimentally, the isolated spirals are kept in place by dielectric plates above and below [11]. Alternatively, thin horizontal plates can be added in the structure in each unit cell and the effects of adding these thin plates (thickness = *3a/128*), where *a* is the pitch of the spiral in the z-direction on the band structure is shown in Fig. 3d. The effect of these thin plates on the polarization gap is very small. From the transmission calculated for circular polarized light in Fig 3(a) and (b), we can deduce that the eigenmode inside the polarization gap is left-hand circular polarized. Here, we adopt the following geometrical definition of the circularly polarized light. At any fixed instant of time, if the tip of the electric field of a plane wave trace out a right-handed helix in space, the plane wave is said to be right-hand circularly polarized and vice versa. This definition will make it easier to understand the physics behind the polarization gap since the polarization gap is produced by the spatial variation of the spiral structures. We note that various chiral structures leading to some form of polarization gaps have also been realized experimentally using vacuum-deposition [12] and liquid crystal structures [13, 14].

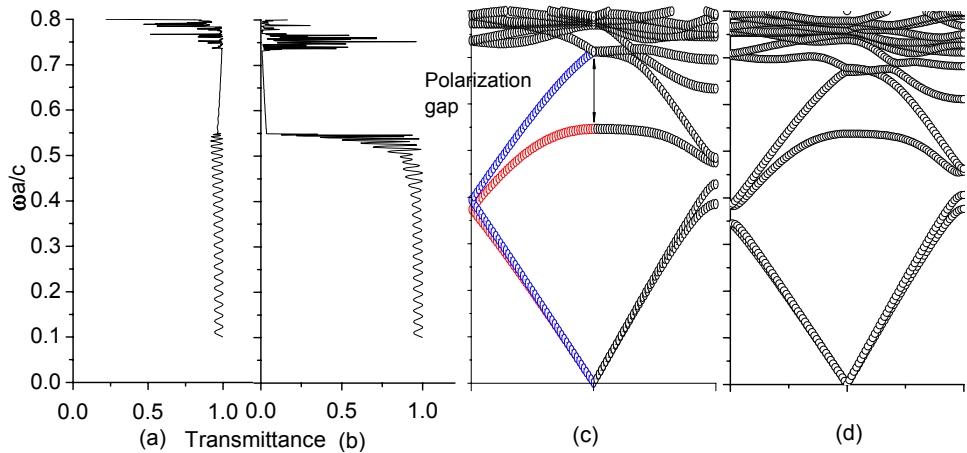

Fig. 3. Transmission spectra and the band structure of the right-handed spiral structure shown in Fig 2. The incident plane wave is left-hand and right-hand circularly polarized in panels (a) and (b) respectively. The band structure is shown in (c). The transmission spectrum is calculated for 16 repeated cells in the z direction. This structure processes a polarization gap, marked with the arrow. In (d), thin plates of dielectric are added in each period in the z direction.

The properties of the eigenmode inside the polarization gap can also be characterized by examining the field patterns. Before making the characterization, the dominant Fourier component of the Bloch wave was identified. Fig. 4 shows the magnitude of the Fourier components (k+G) for the Bloch modes in the ΓA direction. The radius of the open circle is proportional to the magnitude of the plane-wave component in the eigenmode. It shows that the band folding at the Brillouin zone edge is very weak and the wave travels with positive group velocity up to about $\omega a/c=0.7$ since the scattering power of the spirals are not strong enough to make the scattered wave dominant in the ΓA direction. Therefore, it is more instructive to discuss the dispersion in an extended-zone scheme.

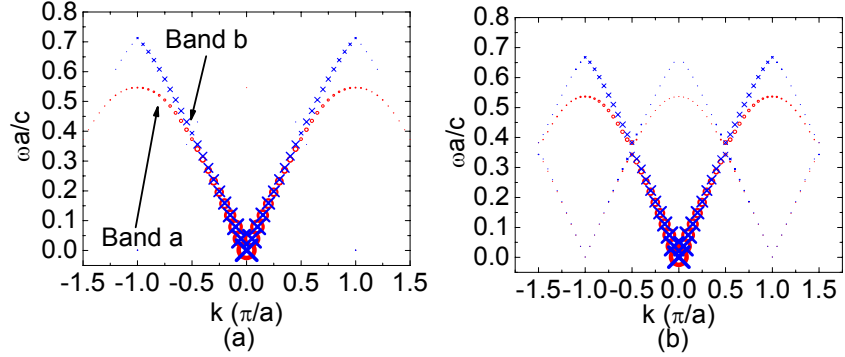

Fig. 4. Band structure along ΓA plotted in the extended zone scheme with the magnitude of the plane-wave components indicated by the radius of the circle. Panel (a) is for the structure with continuous spirals and (b) is for the structure with thin plates inserted in each pitch. Bragg scattering is barely observable in (a). Band "a" and "b" are marked by circles and crosses, respectively.

The eigenmodes are analyzed in two ways. We first consider the ratio $\dfrac{\langle H_x \rangle}{\langle iH_y \rangle}$ where the spatial average $\langle ... \rangle$ is taken inside a unit cell. This method gives a good effective representation for low frequency eigenmodes where the mode structure is simple. The real parts of the ratios are shown in Fig. 5. The imaginary is very small in this frequency range. The results show for the right-handed spiral structure shown in Fig. 2, the band "a" is right-hand, while band "b" is left-hand polarized, and the eigenmode inside the polarization gap is left-hand circularly polarized, consistent with transmission calculations.

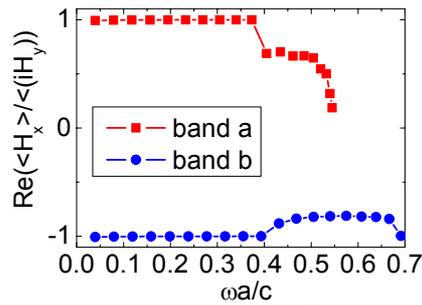

Fig. 5. The ratio between different components of the H-field in the (right-handed) spiral structure of Fig. 2. The ratio is expected to be 1 (-1) for right (left) hand circularly polarized wave.

The second way to do the characterization is by finding the coupling magnitude of the eigenmodes with a plane-wave with well-defined circular polarization. The coupling coefficients are defined as

$$C_\pm(z_0) = |\iint \frac{1}{\sqrt{2}}\{(\hat{x} \pm i\hat{y})^* \cdot \vec{H}(x,y,z_0)\}dxdy|^2, \quad (2)$$

where $\vec{H}(x,y,z_0)$ is the field of an eigenmode in a plane at $z=z_0$. $C_+$ and $C_-$ are respectively the coupling coefficient of left and right-hand polarized plane wave to the normal mode, and the value depends on $z_0$, which is varied to obtain the upper and lower bounds of ratio $\frac{C_-}{C_+}$. If the ratio is very large or very small, the eigenmode is circularly polarized. In the calculation, we considered 128 values of $z_0$. The results are shown in Fig. 6. The ratio is calculated along the band "a" and band "b" and the results are completely consistent with those shown in Fig. 5. The above analyses are also consistent with results obtained from the SMM calculation. As a control calculation, this analysis was also applied to a structure with cylinders of circular cross-section put in a triangular lattice for a control test. The results produced were as expected ($\left|\frac{C_-}{C_+}\right| \sim 1$), i.e., the eigenmodes are linearly polarized for non-chiral structures.

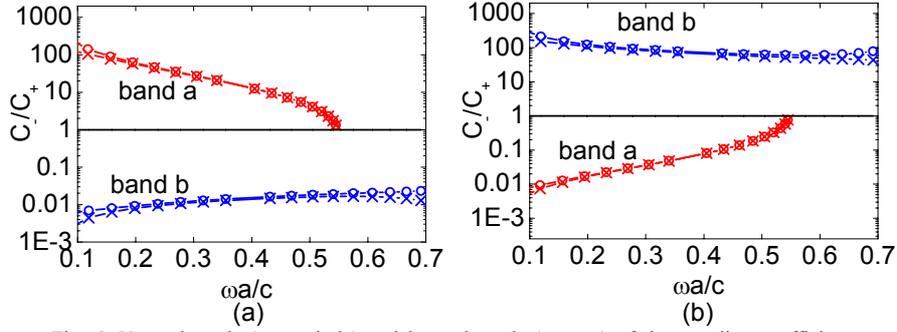

Fig. 6. Upper bounds (open circle) and lower bounds (crosses) of the coupling coefficient ratios (see text) of a circular-polarized plane wave with the eigenmodes for spiral structures. Panels (a) and (b) show the coupling coefficients for if the dielectric spiral is right-handed or left-handed respectively.

The eigenmode characterization shows that the polarization gap for the structure in Fig. 2 will forbid the propagation of the right-hand circularly polarized light. This can be explained by the fact that a right-hand circularly polarized light with a wavelength that matches the pitch of the spiral will match the symmetry of the right-handed spiral structure. As a result, there are two eigenmodes at that wavelength, one with a higher frequency by concentrating its electric field in the air, while the other localize its electric field mainly in the dielectric and has a lower frequency. A polarization gap is then formed. For left circularly polarized light, it is not geometrically possible for the field to "follow" the right-handed spiral, and hence no gap.

From the results in Fig. 3(a), 3(b), 5 and 6, it is expected that if an unpolaried beam is incident to this spiral structure along the ΓA direction, it will be separated into one right-hand circularly polarized beam and one left-hand circularly polarized beam. Inside the polarization gap, the left-hand circularly polarized beam will be transmitted through the spirals and the right-hand one will be reflected. Hence, this structure behaves as a polarization filter.

We now examine the polarization gap as a function of $k_{//}$, the k-vector component parallel to the x-y plane, and results are shown in Fig. 7 along with $k_{//}$ along ΓM and ΓK. The result shows that the polarization gap is fairly robust against incidence angles.

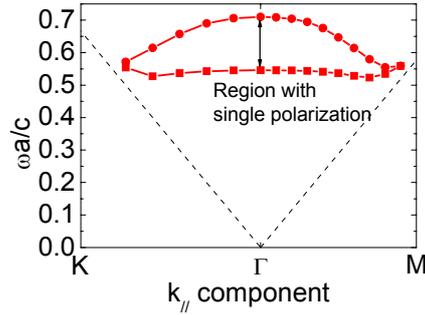

Fig. 7. The polarization gap as a function of $k_{//}$. The area between the two curves is the frequency range in which only a single polarization is allowed inside the spiral structure. The dashed lines represent $k_{//} = \omega/c$.

In summary, we have performed band structure and transmission calculations for dielectric spiral photonic crystals depicted in Fig. 2, and showed that dielectric spiral photonic crystal exhibit significant polarization gaps in which only one circular polarization is allowed. The eigenmodes are analyzed and found to be circularly polarized. In contract to a quarter-wave plate, there is no specific requirement on the thickness of the slab in order to filter out one circular polarization as long as the spiral structure is thick. We suggested laser interference configurations that can achieve such geometries which was demonstrated experimentally [11].

This work is supported by Hong Kong RGC through 600403.